# Optimal Estimation of Capacity and Location of Wind, Solar and Fuel Cell Sources in Distribution Systems Considering Load Changes by Lightning Search Algorithm

Hassan Shokouhandeh
*Electrical and Computer Engineering Faculty*
*Semnan University*
Semnan, Iran
Shokouhandeh@semnan.ac.ir

MahmoodReza Ghaharpour
*Mazandaran regional electric company*
Sari, Iran
mrghaharpour@gmail.com

Hamid Ghobadi Lamouki
*Electrical and Computer Engineering Faculty*
*Islamic Azad University*
Ghaemshahr, Iran
Gh.hamidla@yahoo.com

Yaser Rahmani Pashakolaei
*Electrical Power Distribution Company of Mazandaran*
Sari, Iran
Rahmaniy638@gmail.com

Fatemeh Rahmani
Department of Electrical Engineering
Lamar University
Beaumont, TX, USA
frahmani@lamar.edu

Mahmood Hosseini Imani
Department of Energy
DENERG
Politecnico di Torino
Turin, Italy
mahmood.hosseiniimani@polito.it

*Abstract*— In this paper, estimation of optimal capacity and location of installation of wind, solar and fuel cell sources in distribution systems to reduce loss and improve voltage profile, by considering load changes, is carried out with Lightning Search Algorithm (LSA). Studies have been conducted on the standard IEEE 33 bus power system in two scenarios. In the first scenario, study is done with the assumption that the load remains constant throughout the project period and in the second scenario with the load growth. The results of the simulation, indicated that the puissance of load changes on distributed generations (DGs) placement studies. Also, the results confirm the performance of the proposed algorithm in reducing the objective function.

*Keywords*— *Distribution system - Distributed Generation - Load changes - Lightning Search Algorithm*

## I. INTRODUCTION

Renewable energy sources such as solar and wind have been increasingly used in the electricity industry to reduce fuel costs and thus generate significant profits and clean air growth. This growth has been supported, on the one hand, by fears of a shortage of conventional energy resources and on the other by clean energy production, by interested governments [1]. The dependence of solar and wind power generation on solar radiation and wind speeds has led to lower reliability in renewable systems. To solve this problem, it is suggested to use distributed generation resources that are not affected by environmental parameters. One of these resources that you have been using for the last few years is fuel cells. The performance of fuel cells is very similar to storage batteries except that the reactors and products are not stored and feed the cell continuously. The use of distributed generation sources, as hybrid sources, will increase its efficiency. But the best location and optimal capacity for these resources must be selected. Otherwise, they may be ineffective or may not only improve the network but also make it worse. Therefore, it is appropriate to use meta-algorithms to solve the problem.

In [2] the optimal location of solar and wind power generation sources is determined by intelligent algorithms and the load varies with time; the amount of power generated by the power generation unit is unknown. In [3], a new hybrid method is presented which uses the optimization and distribution of the optimal power. Technical constraints are also used in the optimization process to find the best places to connect DGs. The results of the proposed algorithm are compared with the genetic algorithm, which shows that the proposed method yields a better response. In [4], locating and sizing of solar-wind sources in a distribution network is proposed to achieve minimum losses, increase voltage stability and improve voltage regulation index. In [5] a comparison between the methods of sensitivity of power loss sensitivity, power stability index, and voltage stability index is proposed to determine the optimum location and size of the generating units in the radial distribution network to improve voltage stability margin. Algorithms have been used in [6] to minimize annual energy losses by optimally allocating renewable energy sources. This method is applied to the IEEE 33 standard radial distribution system. The simulation results show that the proposed method is more efficient in minimizing the loss and convergence of the objective function. In [7], a new method of multi-objective optimization algorithm is proposed that aims to determine the optimal location and size of DGs and the contracted price of their generated power. The main objectives of this paper are to reduce the line ohmic losses and improve the voltage profile of the grid. In [8], the optimized method of finding the optimal location of distributed generation sources has been used to simultaneously minimize power losses, improve voltage stability index and voltage deviation of radial distribution network. In [9], a new method for optimizing solar-wind and fuel cell renewables is proposed that uses an intelligent algorithm with operators based on Gaussian probability functions. The proposed method is proved by two examples. The simulation results show that the proposed method can achieve an optimal response in the general search with very low computational time. In the proposed method of reference [10], an integrated approach of loss sensitivity coefficient and voltage stability index is implemented to determine the optimal connection point of distributed generation sources. The simulation results show the performance and effectiveness of the proposed method well. In [11], two methods of sensitivity analysis and optimization algorithms are used. Sensitivity analysis is a systematic technique used to reduce the search space and to arrive at a correct solution for locating capacitors.

In this paper, optimal location and capacity of solar and wind sources and also fuel cell with regard to load changes is studied. The optimization process will aim to reduce ohmic losses and improve the voltage profile in all buses along the feeder. For this purpose, the modified version of lightning search algorithm is proposed.

## II. THE SYSTEM UNDER STUDY

In Fig. 1, the single-line diagram of the under-study IEEE 33 bus system is shown. This feeder has a radial structure. Bus number one is selected as slack bus.

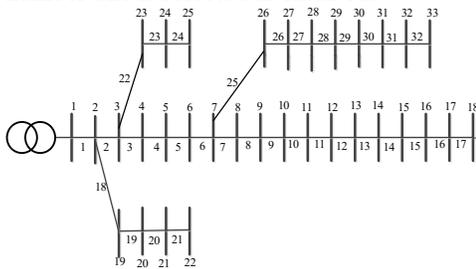

Fig. 1. Single line diagram of IEEE 33 bus system

In Fig. 2, the load profile is shown during the study period.

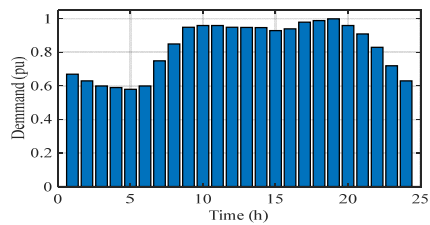

Fig. 2. Network load profiles during the study [12]

The minimum load required is less than 0.58 pu at 5 o'clock and the peak load at 19 o'clock is assumed to be 1 pu. Fig. 3 shows the ohmic losses of the lines due to the feeder load changes in the 24 h study.

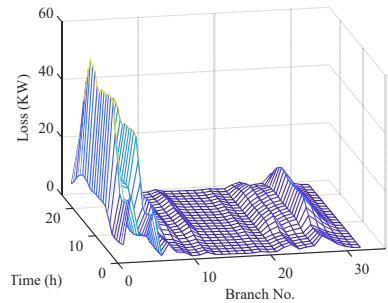

Fig. 3. The 33-bus system line losses in 24 hours

The total active losses in this feeder is estimated about 1618 kW for the 24 hour. Also, the maximum loss is related to line number 2, which occurred at 7 pm. The voltage deviation index of the system is shown in Fig. 4. The value of the voltage deviation index in bus 18 is lower than in other buses. So that at 7 pm, the voltage range in this bus reaches about 0.86 pu.

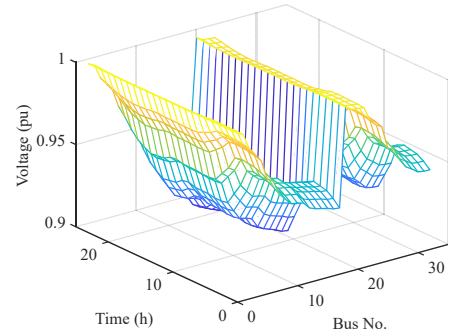

Fig. 4. The voltage profile in the 33 bus system

Generating capacity of wind and solar sources depends on the intensity of solar radiation and wind speed. Fig.5 shows daily wind speeds and solar irradiations.

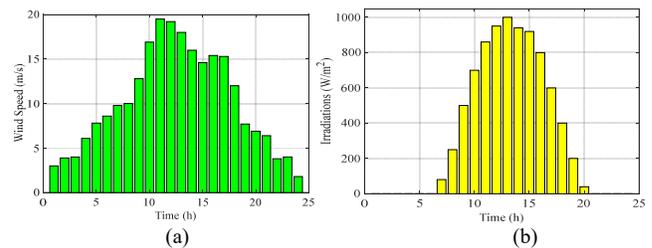

Fig. 5. a. Daily radiation intensity b. Daily Wind Speed [12]

According to the Fig. 5, it is possible to get solar energy in the grid from 6 to 21 hours. The wind speed also varies from 2.5 m/s to 19.7 m/s.

## III. OBJECTIVE FUNCTION AND CONSTRAINTS

In order to determine the optimal capacity of DGs, considering the load changes, an appropriate objective function must be provided. In the this study, loss and voltage deviations are selected as the main objectives. The ohmic losses of the lines can be calculated as equation (1). The loss function is the sum of the ohmic losses of all the lines [13-14].

$$\text{Loss} = \sum_{t=1}^{24} \sum_{i=1}^{N_b} R_i I_i(t)^2 \quad (1)$$

In the above equation, *Ri* and *Ii* are the values of line resistance and current of the line *i*, respectively, $N_b$ is the number of lines in the feeder. The voltage deviation index is also calculated by equation (2) [13].

$$V_{\text{Profile}} = \sum_{t=1}^{24} \sum_{i=1}^{n} |v_i(t) - v_{\text{nom}}| \quad (2)$$

In the above equation *Vi* is the bus voltage (pu), $V_{nom}$ is the rated voltage in pu and n is the number of buses.

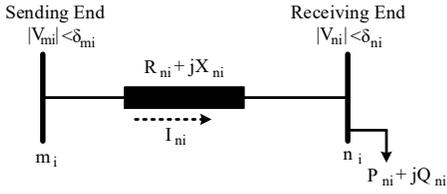

Fig. 6. Part of the radial feeder of the distribution system

Because the lightning search algorithm can optimized one objective function, so, loss and voltage deviation functions must be converted to one objective function by weighted coefficients. Thus the main objective function will be in the form of the following equation.

$$\text{OBJ} = \omega_1 \text{Loss} + \omega_2 V_{\text{Profile}} \quad (3)$$

By using the weight coefficients $\omega_1$ and $\omega_2$, the two objective functions are weighted to have an equal effect on the final value of the objective function. Connecting the distributed generation sources to the feeder in question should be done in such a way that the normal operation of the network does not change. Therefore, constraints should be considered to solve the optimization problem by algorithms. The most important limitations in locating and determining the capacity of DGs in the distribution feeder are [13]:

The amount of active and reactive power produced and consumed should be equal after the installation of DGs in the feeder.

$$P_{\text{grid}} + P_{\text{DG}} = P_{\text{Demmand}} + P_{\text{Loss}} \quad (4)$$
$$Q_{\text{grid}} + Q_{\text{DG}} = Q_{\text{Demmand}} + Q_{\text{Loss}} \quad (5)$$

In the above equations, $P_{grid}$ and $Q_{grid}$ are active and reactive power interchange with main grid, $P_{DG}$ and $Q_{DG}$ are active and reactive power generated by DG units, $P_{Demmand}$ and $Q_{Demmand}$ are active and reactive demand by consumers and $P_{Loss}$ and $Q_{Loss}$ are active and reactive losses.

The second optimization limitation is that; all the buses voltage should be within the permitted range [15].

$$|v_i^{\min}| \le |v_i| \le |v_i^{\max}| \quad (6)$$

The current amplitude in all branches should also not exceed the permitted capacity of the line after compensation.

$$|I_i| \le |I_i^{\max}| \quad (7)$$

In addition to network constraints, there are also limitations to the capacity and number of distributed generation resources, which can have technical or economic reasons.

$$P_{DG}^{M\,in} \le P_{DG} \le P_{DG}^{M\,ax} \quad (8)$$
$$Q_{DG}^{Min} \le Q_{DG} \le Q_{DG}^{Max} \quad (9)$$
$$N_{DG} \le N_{DG}^{M\,ax} \quad (10)$$

## IV. LIGHTNING SEARCH ALGORITHM

The Lightning Search Algorithm is a meta-heuristic optimization algorithm modelled on the natural phenomenon of lightning in 2015. This algorithm uses the propagating branch propagation mechanism in lightning. In the LSA algorithm, faster particles, called projectiles, are considered as binary tree structure, as a progressive branch. Progressive branches formed by projectiles determine the initial population size. In fact, projectiles offer random solutions to the problem by lightning search algorithm. A projectile loses its kinetic energy through movement in the atmosphere and by tensile collisions with air molecules. The velocity of the projectile and its kinetic energy are calculated as follows [15-16].

$$v_p = \left[1 - (1/\sqrt{1-(v_0/c)^2} - sF_i/mc^2)^{-2}\right]^{-0.5} \quad (11)$$

$$E_p = \left((1/\sqrt{1-(v_p/c)^2}) - 1\right) mc^2 \quad (12)$$

In the above equations, $v_0$ is the initial velocity of the projectile, $F_i$ is the constant ionization rate, *m* the projectile mass and *c* is the light velocity. The velocity and kinetic energy depend on the position of the progressive branch and the mass of the projectile. Therefore, when the mass is small or long distances, then the energy projectile has less potential for ionization. At the same time, the exploration power and capacity of this algorithm can be controlled by the relative energies of the progressive branches. Due to the formation of a projectile that exits in a random direction during the lightning transition, the tip of the arrow forms at the same initial stage. Therefore, it can be modelled as a random number that uses the probability of standard uniform distribution over the search space. The probability density function is defined as follows.

$$f(x^T) = \begin{cases} 1/b - a & a \le x^T \le b \\ 0 & x^T < a \,||\, x^T > b \end{cases} \quad (13)$$

In this respect, $x^T$ is a random number that generates the initial energy of the progressive branch, a and b constitute the upper and lower limits of the search space, respectively. For a population consisting of N progressive branches, N random particles are required. As the progressive branches evolve, they must shift their position by ionizing the parts in the vicinity of the best response in the next step. The location of space projectiles in the next step can be roughly considered as a random number generated by exponential distribution.

$$f(x^s) = \begin{cases} \dfrac{1}{\mu} e^{\frac{-x^s}{\mu}} & x^s \ge 0 \\ 0 & x^s < 0 \end{cases} \quad (14)$$

μ is the coefficient of formation that can control the projectile's location. The location of the projectile can be calculated from the following relation:

$$P_{i\_new}^{S} = P_{i}^{S} \pm \text{ExpRand}(\mu_i) \quad (15)$$

In the above equation ExpRand is an exponential random number that if negative, then the generated random number must be negative. The new location does not guarantee progressive branch propagation or channel formation unless the projectile energy value is higher than the progressive branch to find a good response. If the right solution is found in the next step, the progressive directory will be moved to the new location and updated. Otherwise, the values will remain constant until the next step. If it goes on, it will be the guiding projectile leader along the way. Progressive branch projectiles that have reached the nearest location to the ground will not have the potential to ionize large sections against the tip. Therefore, the projectile is modeled as a random number from the normal probability distribution function expressed in Equation (16).

$$f(x^L) = \frac{1}{\sigma\sqrt{2\pi}} e^{\frac{-(x^L-\mu)^2}{2\sigma^2}} \quad (16)$$

The standard deviation for the guided projectile is reduced exponentially until it is transported to the ground. In other words, he finds the best solution. So the new location is displayed in the following relationship form:

$$P_{i\_new}^{L} = P_{i}^{L} \pm \text{NormRand}(\mu_L, \sigma_L) \quad (17)$$

In the above case NormRand is a random number generated by the normal probability density function. The new location does not guarantee the propagation of the progressive branch unless the energy is increased beyond the reach of the solution to the problem.

## V. SIMULATION RESULTS AND ITS ANALYSIS

Choosing the optimal capacity and location for installing DGs greatly impacts on their performance. The simulations are done in two scenarios, in the first scenario, optimal location and sizing of the DGs in the system without load changes and in the second scenario, considering the load changes. For this purpose, LSA algorithm, the Particle Swarm Optimization (PSO) algorithm is used, the parameters of the algorithms are presented in Table (I).

TABLE I.  THE PARAMETERS OF PSO AND LSA ALGORITHM

| LSA | Pop | Iter | $\gamma$ | $\rho$ | | |
|---|---|---|---|---|---|---|
|  | 100 | 35 | 0.2 | 0.8 | | |
| PSO | Pop | Iter | C1=C2 | $V_{min}$ | $V_{max}$ | $\Omega$ |
|  | 100 | 35 | 2 | 0.4 | 0.9 | 0.78 |

The population and iterative values for both algorithms in solving the optimization problem are chosen equal to compare the performance of the algorithms fairly.

### A. First Scenario (Excluding Load Changes)

In this section, the placement and sizing of solar and wind sources and fuel cell are performed without considering load variations and assuming the load remains at nominal value. It is also assumed that it is only possible to install a generator of any type with a limited capacity along the feeder. After performing the optimization, the convergence process of the two algorithms in minimizing the objective function is shown in Fig. 7. Final value of the objective function for the LSA algorithm is 1.48pu obtained after 29 iterations. However, final value of the PSO algorithm is 1.55pu.

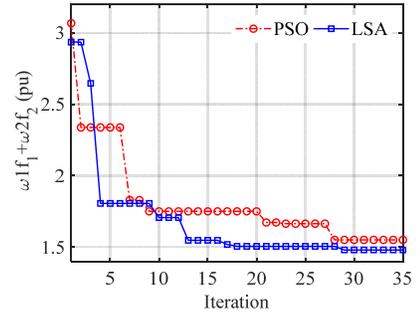

Fig. 7. Convergence of LSA and PSO algorithms in the first scenario

Figure 8 shows the values of the loss and voltage functions after the installation of the distirburted generations.

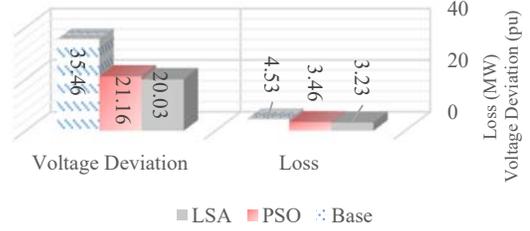

Fig. 8. Simulation results in the first scenario

The amount of loss and voltage deviation index in these conditions after optimization by LSA algorithm equals to 3.23 MW and 20.03pu and if optimized by PSO algorithm the values of these two indices are equal to 3.46MW and 21.16pu. Table II shows the optimal location and capacity of the DGs.

TABLE II.  OPTIMAL LOCATION AND CAPACITY OF DISTRIBURTED GENERATIONS IN THE FIRST SCENARIO

|  | PSO | | LSA | |
|---|---|---|---|---|
|  | Location | Size | Location | Size |
| PV | 32 | 2150 | 32 | 2210 |
| WT | 18 | 1940 | 16 | 1730 |
| FC | 8 | 930 | 6 | 870 |

According to the results recorded in Table II, after optimally estimating the size and location of the distributed generation sources by the PSO algorithm, the 32, 18 and 8 buses are chosen for the installation of the PV, WT and FC with capacities of 2150 kW, 1940 kW and 930 kW respectively. But if using the proposed LSA algorithm to locate and determine the capacity of DGs in the system, bus 32 for installation of a 2210 kW PV panels, bus 16 for 1730 kW WT and finally bus 6 is selected to install 870 kW FCs. The loss of the lines after installing the DGs along the feeder are shown in Fig. 9. Accordingly, the ohmic losses in all the transmission lines have fallen sharply after the installation of distributed generation sources.

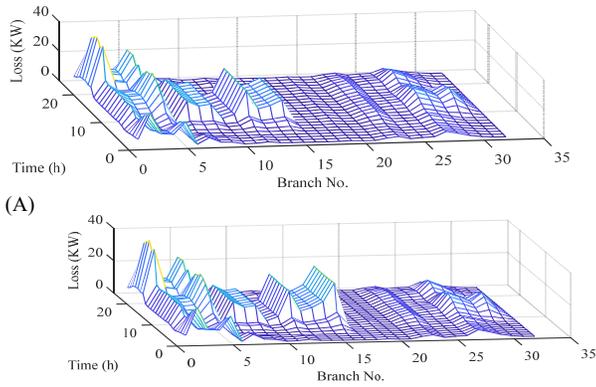

(A)

(B)

Fig. 9. The lines losses in the first scenario (A. LSA B. PSO)

The amount of feeder losses studied after installing DGs compared to pre-compensation conditions decreased by about 23.6% when using the PSO algorithm and by 28.7% when using the proposed LSA algorithm. On the one hand, the installation of DGs along the feeder has improved the voltage profile of the grid. The voltage deviation function after installation of the DGs at the locations proposed by the LSA algorithm is about 43.6% and if these sources are installed at the locations and capacities suggested by the PSO algorithm, about 43.3% is improved. In Fig. 10, the voltage amplitude of all buses is shown. According to the obtained values, the maximum voltage deviation range is less than 0.09 pu after installation of the DGs.

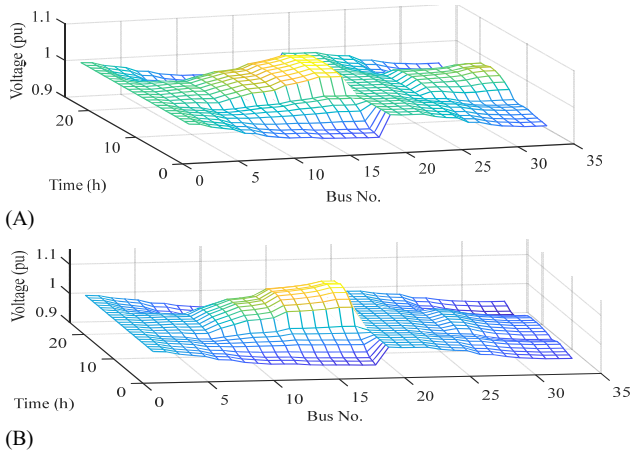

(A)

(B)

Fig. 10. Voltage profiles in the first scenario(A. LSA B. PSO)

### B. Second scenario (Load changes)

To minimize losses and voltage deviations in the feeder, growth and load changes must be considered during the study period. Because the load changes of the system will cause significant changes in the loss rate as well as the voltage profile of the grid. But the important issue is to choose the right place to install distributed generation and determine their optimal capacity. Because if the location or capacity of each of these sources is not selected correctly, the desired results will not be achieved and it is possible that the voltage range of the chains will be out of range. In this scenario, the PSO and LSA algorithms are used to solve the optimization problem. The convergence process of the responses of the two algorithms in different iterations is shown in Fig. 11.

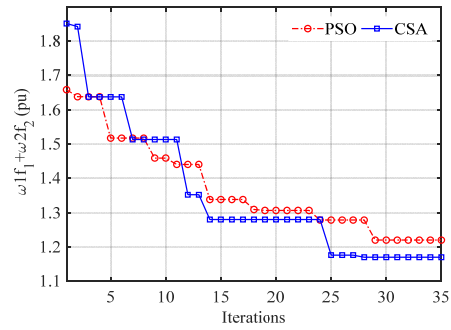

Fig. 11. The convergence process of lightning and particle swarm search algorithms in the second scenario

The final value of the objective function after optimization by lightning search algorithm is equal 1.17 pu while the optimal value obtained by the PSO algorithm is 1.22 pu. The final value of the objective function is less than the first scenario, which means lower loss and voltage deviation. For a more detailed analysis of the simulation results in the second scenario, the values of loss and voltage deviation function are calculated separately and are shown in Fig.12.

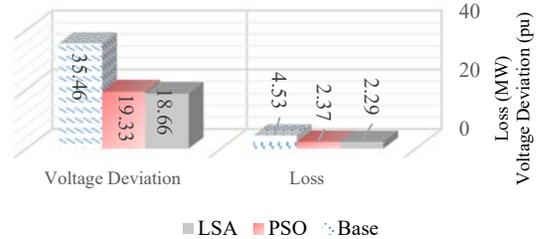

Fig. 12. Simulation results in the second scenario

The amount of loss and voltage function in the second scenario after optimization by LSA equals 2.29 MW and 18.66 pu and if optimized by PSO equals 2.37 MW and 19.33 pu. In Table III, the optimal location and capacity of the PV, WT and FC are given.

TABLE III. OPTIMAL LOCATION AND CAPACITY OF DGS IN THE SECOND SCENARIO

|    | PSO      |      | LSA      |      |
|----|----------|------|----------|------|
|    | Location | Size | Location | Size |
| PV | 12       | 1645 | 13       | 1465 |
| WT | 30       | 1725 | 29       | 1835 |
| FC | 19       | 1315 | 3        | 1260 |

The locations proposed by the PSO algorithm for installing solar, wind and fuel cell sources are 12, 30 and 19, respectively, with capacities of 1645 kW, 1725 kW and 1315 kW, respectively. Whereas the proposed LSA algorithm proposed installing PV and WT with capacities of 1465 kW and 1835 kW in bus 13 and 29 and installation of FC with capacity of 1260 kW in bus 3. The ohmic losses of lines in the standard 33 bus feeder, after locating and determining the optimum capacity of the distributed generation resources are shown in Fig. 13.

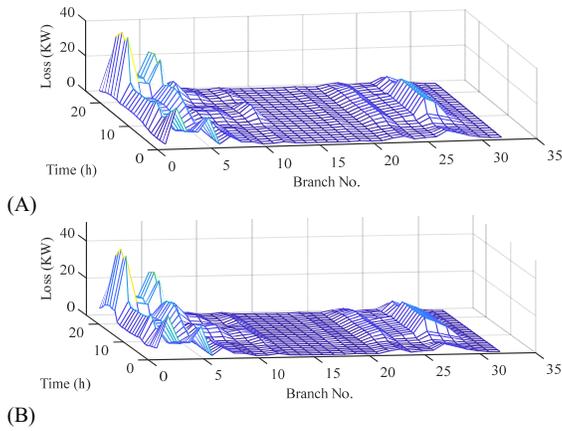

Fig. 13. Lines losses in the second scenario(A. LSA B. PSO)

The loss of most lines is lower when installing DGs at the locations and capacities specified by the proposed LSA algorithm. Under these conditions, after the installation of distributed generation sources along the feeder, the ohmic losses are reduced by 49.01% when using the PSO algorithm and by 47.7% when using the LSA algorithm. In order to investigate the modified amount of voltage profiles along the feeder after the installation of the distributed products, the voltage amplitude for all the buses are shown in Fig. 14.

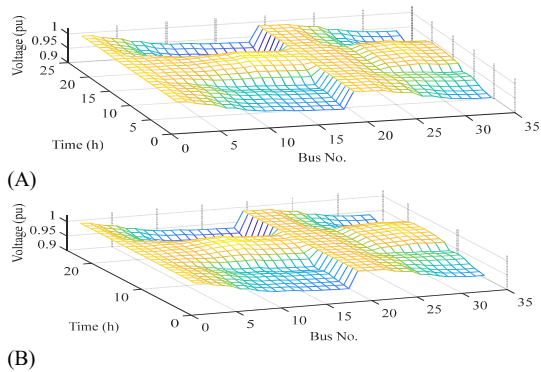

Fig. 14. Voltage profile after using DGs for second scenario (A. LSA B. PSO)

As shown in the figures above, the presence of distributed generation sources has led to the improvement of the voltage profile in all the buses. The voltage deviation function in this feeder improved about 45.5% after optimization by PSO algorithm and about 47.4% after optimization by LSA algorithm.

## VI. CONCLUTION

By comparing the results, it can be deduced that the presence of distributed generation sources is significantly effective in reducing losses. But to minimize losses, load changes must be considered in the studies. In the first scenario, the reduction in casualties is assumed to be about 28.7%, while the reduction in ohmic loss is more than 47.7% when considering load changes. Just as the presence of distributed generations can reduce feeder ohmic losses, it can also improve the power quality of the grid and improve the voltage profile. So that after installing the distributed output sources the voltage amplitude in all the trays increased. After installing the distributed generations with constant load assumption, the voltage deviation function is improved to 43.6%. % Reaches.